# Exploring Scientific Contributions through Citation Context and Division of Labor

Liyue Chen[1], Jielan Ding[1,2*], Donghuan Song[1], Zihao, Qu[1]

[1]National Science Library, Chinese Academy of Sciences; Beijing, 100190, China.
[2]Department of Information Resources Management, University of Chinese Academy of Sciences; Beijing, 100190, China

**Correspondence to:**

Jielan Ding

National Science Library, Chinese Academy of Sciences; Beijing, 100190, China

Department of Information Resources Management, University of Chinese Academy of Sciences; Beijing, 100190, China

**Email:**    dingjielan@mail.las.ac.cn

**Abstract**

Scientific contributions are a direct reflection of a research paper's value, illustrating its impact on existing theories or practices. Existing measurement methods assess contributions based on the authors' perceived or self-identified contributions, while the actual contributions made by the papers are rarely investigated. This study measures the actual contributions of papers published in *Nature* and *Science* using 1.53 million citation contexts from citing literature and explores the impact pattern of division of labor (DOL) inputs on the actual contributions of papers from an input-output perspective. Results show that experimental contributions are predominant, contrasting with the theoretical and methodological contributions self-identified by authors. This highlights a notable discrepancy between actual contributions and authors' self-perceptions, indicating an "ideal bias." There is no significant correlation between the overall labor input pattern and the actual contribution pattern of papers, but a positive correlation appears between input and output for specific types of scientific contributions, reflecting a "more effort, more gain" effect. Different types of DOL input in papers exhibit a notable co-occurrence trend. However, once the paper reaches the dissemination stage, the co-occurrence of different types of actual contributions becomes weaker, indicating that a paper's contributions are often focused on a single type.

# 1. Introduction

Scientific contributions directly reflect the value of a paper, showcasing its inspiration, promotion, and enhancement of existing theories or practices. Exploring scientific contributions is crucial for understanding the research content and innovative value of scientific papers. The scientific contribution of a paper is a complex concept involving aspects such as types and strength of contributions. Contribution type refers to the specific aspects of the research process that a paper provides new knowledge or insights into(Cai et al., 2023). Common types of contributions include theoretical framework construction, methodological development, and empirical findings. It helps researchers understand the main value and novelty of the study. While contribution strength emphasizes the impact of contributions on academic and social development, and specific dimensions include plausibility, originality, scientific value, and social value(Lamont, 2009; Polanyi, 1962).

Scientific contribution measurement methods can be divided into two aspects. On one hand, citation-based methods are often used as a proxy for measuring scientific contributions or research quality (Caon et al., 2020; Cole & Cole, 1971), as certain features of a paper's contribution are correlated to the distribution and variation of citations (Aksnes et al., 2019). These metrics essentially measure citation impact and, to a certain extent, reflect the contribution strength of papers. However, existing studies have shown that these metrics have low reliability when measuring the scientific contribution of individual papers (Aksnes, 2006; Aksnes et al., 2023), and they provide limited insight into the types and patterns of scientific contributions. On the other hand, currently, many studies extract contribution arguments by identifying typical expressions in the full text of papers to measure their main types of scientific contributions(Chao et al., 2023; Chen et al., 2022). The results of such measurement methods can be referred to as expected contributions or self-perceived contributions since they are derived from the authors' firsthand knowledge of their publications and research fields, which can elucidate the significance of their works to scientific development(Aksnes, 2006).

However, authors' perceptions on contributions may differ from those of their peers. Therefore, this study aims to measure the contributions of papers to other research in the context of scholarly communication, which we term actual contributions or peer-absorbed contributions. The rapid development of open science has enabled extensive access to machine-readable full-text of papers, offering rich citation context data that presents new opportunities to measure the scientific contributions of individual papers and explore their fine-grained characteristics. Our measurement approach is based on Merton's norms of science, where researchers acknowledge or credit the contributions of the cited research through citation behavior(Merton, 1973). Furthermore, from the input-output relationship, this study also explores the influencing factors related to the input aspects that affect the actual contributions of a paper, with a focus on a typical aspect of input for papers known as division of labor (DOL) (Haeussler & Sauermann, 2020; Lee et al., 2015). This labor input also reflects, to some extent, the efforts of a paper to advance scientific research, and can thus be seen as a form of input contribution, which we term input effort of DOL.

Overall, different from the self-identified contributions measured by those existing research which often reflect the authors' expectations, this study explores the actual contributions of a paper through its utilities in the citation context of citing papers. Thus, the research questions are as follows:

(1) What are the actual contribution types of a paper, i.e., the contribution types a paper makes to another research?

(2) From the perspective of the input-output relationship in scientific activities, what are the input effort of DOL associated with the actual contributions of a paper?

(3) Is there a coupling or isolation relationship between different types of scientific contributions of a paper?

To answer these questions, we raise a taxonomy for actual scientific contributions and design a method for identifying the types of scientific contributions using large language model technology and citation context. Moreover, we propose a scheme for associating DOL with scientific contributions, aiming to reveal the relationship between actual contributions and input efforts of DOL. This involves correlating and mapping labor division with corresponding types of scientific contributions, followed by a correlation analysis between different types of scientific contributions from the perspective of both input efforts of DOL and actual contributions.

## 2. Related Work

### 2.1 Scientific Contributions of Papers

Scientific contribution is a focal point in the research evaluation system, closely linked to critical scientific decisions such as funding allocation, career assessment, and awards selection. Traditional methods for assessing scientific contributions primarily rely on peer review to evaluate the types and strength of contributions within a limited scope of research papers or scientists. With the rapid development of natural language processing and artificial intelligence technologies, some studies have begun exploring the automatic identification of scientific contribution types in academic papers.

Clearly defining the specific types of scientific contributions is fundamental to designing automatic identification methods. Existing research mainly delineates contribution types by observing the textual features describing scientific contributions within academic publications. Early studies approached the classification of scientific contribution types from a broad perspective, categorizing them into general types, primarily including empirical, theoretical, method, and other types (Aksnes, 2006; Porter et al., 1988). In recent years, some studies have focused on the field of computer science and proposed fine-grained taxonomies. Chen et al. (2022) introduced a six-category classification including theory proposal, model creation and optimization, and algorithm development. Chao et al. (2023) proposed a framework that includes eight contribution types along with fifteen corresponding contribution arguments. D'Souza and Auer (2020) categorized contributions into ten types based on core information units in research activities, such as research problem, methods, results, and code.

Based on various scientific contribution taxonomies, some studies have proposed automatic identification methods for scientific contribution types of papers. One type of research mainly uses machine learning models to achieve automatic recognition. Cai et al. (2023) proposed a research contribution recognition model enhanced by semantic role annotation based on SciBERT. Similarly, Liu et al. (2023) divided scientific contributions into four types: approach, analysis, result, topic or resource. They fine-tuned SciBERT by manually annotating contribution type data, and the model finally achieved a Micro-F1 result of 90.5. In addition, some studies have undertaken similar tasks of assessing scientific contributions by automatically identifying innovation-related sentences within papers(Wang et al., 2018; Zhang & Chen, 2024).

Some studies have used knowledge graphs of contributions to represent the scientific contribution types of papers. Auer et al. (2018) constructed an open research knowledge graph (ORKG), where each paper is represented by its fundamental contribution types and their contribution values. Vogt et al. (2020) modeled the scientific contributions of papers using knowledge graph cells, with ontology concepts including objective, result, method, material, activity, and agent. Gupta et al. (2024) proposed a method to inject factual knowledge from scientific knowledge graphs into pre-trained models to identify significant scientific contributions in papers. Additionally, recent studies have explored the use of prompt engineering and generative large language models to identify types of scientific contributions in small-scale paper abstracts, achieving an accuracy of over 80% (Cao et al., 2024).

**2.2 Division of Labor in Research Papers**

The division of labor in academic papers is fundamental to effectively conducting scientific research. Research activities, from conceptual design and data collection to analysis and summarization, require specialized collaboration among researchers. Explicit roles and contributions of team members in research activities helps establish a fair contribution evaluation mechanism, optimize the organization and resource allocation of research teams, and thus promote innovative breakthroughs in science.

Different academic groups define DOL according to their research objectives and management needs. Several organizations have released authoritative frameworks for DOL, including the Contributor Roles Taxonomy (CRediT) and the ICMJE recommendations. CRediT has established 14 DOL roles and is widely used as a reference template for author contribution statements in journal publications. Some journals developed variants of these DOL frameworks, such as *Nature*, *Science* and *PNAS*. Additionally, researchers have refined existing frameworks from the perspective of quantifying contribution levels. Allen et al. (2014) analyzed author contribution statements in journals to categorize author contribution activities into 14 types. Larivière et al. (2016) used author contribution statement data from *PLOS* papers to statistically classify six DOL types. Cai and Han (2020) proposed a taxonomy with seven types using a similar approach. De Souza et al. (2022) mapped authors' specific DOL into theory contribution and methodology/Logistics contribution based on author contribution statements from three medical journals. Herz et al. (2020) noted that standardized DOL taxonomy can reduce the time cost for authors and readers, facilitate accountability and contribution measurement, and decrease the tendency of authors to overestimate their contribution to scientific work.

DOL in academic papers are primarily identified through manual annotation and automatic extraction using machine learning models. Walsh and Lee (2015) and Clement (2014) categorized co-author roles through contribution questionnaires and interviews with paper authors. Sauermann and Haeussler (2017) manually coded and annotated semi-structured author contribution texts from *PLOS ONE* to determine the authors' DOL roles. Currently, automatic methods for identifying DOL have rapidly developed. Rule-based methods define rule libraries based on the semantic structure of author contribution statements and match unstructured text content with these libraries to determine the DOL of authors (Cai & Han, 2020; Larivière et al., 2021). Machine learning-based methods address more complex free-text author contributions by training models such as SVM and random forests to automatically classify contribution activities (Lin et al., 2023; Xu et al., 2022).

## 3. Data and Methods

### 3.1 Data

Our study uses research articles published in *Nature* and *Science* between 2011 and 2023 to measure scientific contributions. The research initially obtained the metadata of 70,710 papers from Web of Science (WoS), and by collecting data from the official journal websites, identified 8,592 articles of the Article type (including Research Articles and Special Issue Research Articles). From these, 7,408 articles containing author contribution statements were selected. Subsequently, using the DOI information of these papers, citation context data from the citing papers was retrieved through API calls from the Scite database. We then retrieved citation context data from citing papers through Scite database by searching DOIs for these papers. We replaced the position of the cited reference in these citation contexts with '[target cited reference]' to facilitate subsequent target localization and automatic extraction. Ultimately, 7,041 papers were identified as having both author contribution (AC) and citation context data, with a total of 1,534,467 citation context corresponding to these papers.

Basic descriptive data on two journals are provided in Table 1. We considered the following issues in constructing this dataset. Firstly, from the citation context dimension, the average citation frequency per paper in the dataset is 117.37, and the average frequency of mentions within the text is 217.93. This indicates that papers published in two prestigious journals have more extensive citation context data, making the corresponding statistics on scientific contributions more representative. Secondly, from the dimension of DOL, these two journals have been pioneers in requiring papers to provide author contribution statements at the publication stage. From 2020 to 2023, approximately one-quarter of papers already have author contribution statements (Lin et al., 2023). The related information is relatively standardized and complete, allowing it to reflect the DOL in papers to a certain extent.

**Table1** Basic descriptive data on Nature and Science journal publications

|  | **Nature** | **Science** | **Nature & Science** |
|---|---:|---:|---:|
| N. publications | 5,331 | 1,710 | 7,041 |
| %. with AC data[*] | 33.02% | 14.88% | 24.85% |
| Avg. Citations | 126.7 | 88.28 | 117.37 |
| Avg. Mentions | 234.44 | 166.47 | 217.93 |

\* The percentage data mainly refers to the proportion of research papers with author contribution statements to all research papers in 2020-2023.

## 3.2 Methods

Figure 1 shows the scheme for measuring and analyzing the scientific contribution of papers. Our scheme includes steps such as measuring the actual contribution of papers, measuring the input efforts of DOL in papers, and analyzing the correlation between the two.

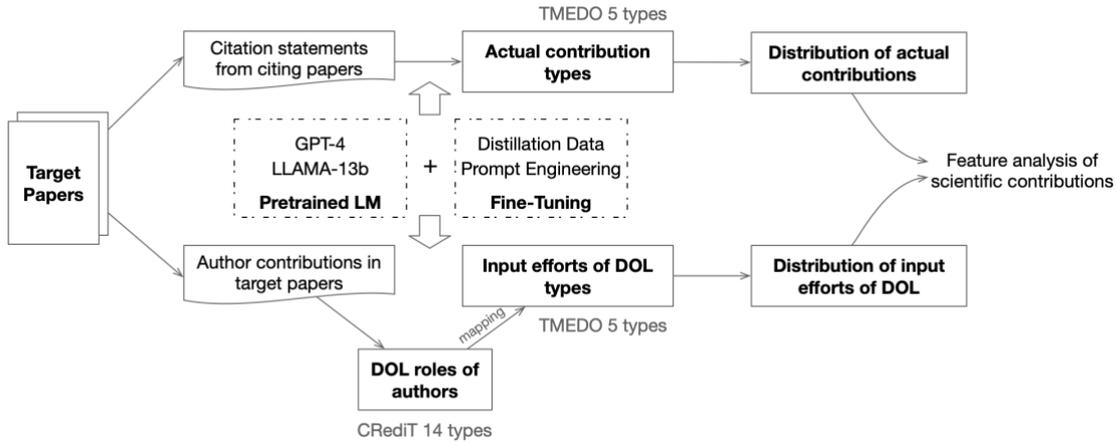

**Figure 1.** Paper scientific contribution measurement scheme

*Measurement of Actual Contributions*

To measure the output or actual contribution of papers, we first need to determine the types of actual contributions. According to existing scientific contribution taxonomies, scientific contribution emphasizes the role of research work in the lifecycle of scientific activities. Additionally, since the target papers we analyze spans multiple disciplines, the taxonomy's applicability across different fields must also be considered. Hence, we proposed a scientific contribution taxonomy comprising five types (abbreviated as TMEDO), with each type and its description presented in Table 2.

**Table 2** Taxonomy of scientific contributions and their descriptions

| Scientific contribution types | Description |
|---|---|
| Theoretical | The citing paper draws upon theories, conceptual models, ideas, or viewpoints proposed by the target cited reference |
| Methodological | The citing paper employs or mentions experimental methods, methodological parameters, or frameworks designed by the target cited reference |
| Experimental | The citing paper utilizes, assesses or discusses the experimental findings or results derived from the target cited reference |
| Data-based | The citing paper uses datasets, sample collections, or other experimental resources constructed by the target cited reference |
| Other | Other contributions of the target cited reference to the citing paper, or contributions that cannot be categorized into the previous four types |

The study further pursued the automatic identification of actual contributions. We initially attempted to identify contribution types using various large-scale language models and prompt engineering (see Figure S1 for the prompt scheme), but found the recognition effect to be relatively limited. We then proposed a recognition optimization scheme, the main idea of which is the refinement of open-source LLMs using a self-learning strategy. During the data annotation phase: 1) We invited three researchers to annotate the scientific contribution types (two for annotation and one for review) and a labeled dataset of 500 citation contexts was constructed ultimately. 2) To efficiently expand the labeled dataset, we designed prompt schemes and used GPT-4-0613 for automatic contribution annotation. Using the small labeled set, we verified that the accuracy of the model reached 0.87. 3) 11,250 labeled data points were automatically generated as the final training dataset, and 1,120 labeled data points, combining manual and automatic annotation, were used as the final test dataset. During the model training phase, we selected pre-trained models such as BERT and open-source large language models like LLaMA and Gemma. As shown in Table 3, after using QLoRA to fine-tune each model and comparing their performance, we found that the fine-tuned LLaMA2-13B model was the most suitable for the automatic identification of scientific contributions in this study, achieving an F1-score of 0.94 (see Table S1 for detailed evaluation results).

**Table 3** Types of scientific contributions and their descriptions

| Models | Accuracy | Precision | Recall | F1-score |
|---|---|---|---|---|
| **LLaMA2-13B** | **0.94** | **0.95** | **0.94** | **0.94** |
| Bert | 0.86 | 0.89 | 0.86 | 0.83 |
| Gemma-7B | 0.48 | 0.52 | 0.48 | 0.44 |
| Gemma-2B | 0.24 | 0.41 | 0.24 | 0.14 |

After identifying the actual contribution types of the target papers, the study further calculated the distribution of output contributions for each paper. Here, we primarily used full counting for calculation, meaning that for a given target paper, each citation context is counted as 1 in the contribution statistics. Since the frequency of mentions within the citing literature reflects the contribution degree of the target paper to some extent(Herlach, 1978; Tang & Safer, 2008), and each mention may represent a different type of contribution, this counting method accounts for both the types and degrees of contributions, ensuring the completeness of the measurement. By using this method, we ultimately obtain the proportion of citation context for each paper across the five contribution types.

*Measurement of Input Efforts of DOL*

Based on our raw data collection of papers, it was found that 7,408 of these papers contained author contribution statements, including 863 structured and 6,545 unstructured paragraphs. To facilitate the extraction of author contribution information paragraphs, this study adopted a data processing logic involving intelligent sentence segmentation, author contribution information alignment, and DOL identification.

Firstly, an intelligent sentence segmentation program was developed to split the contribution description paragraphs into sentence-level content, thereby obtaining sentence-level descriptions that include both the 'author name' and the 'author's contribution.' After manually removing some irrelevant descriptions lacking author or contribution information, 54,318 valuable sentences were retained.

Secondly, an author contribution information alignment experiment was conducted. 500 papers were randomly selected, and their valuable sentences were used as labeled data. A combination of GPT-3.5 and manual correction was employed to create a training set of 3,095 entries and a test set of 714 entries. Both the training and test sets were formatted as Q&A, where the questions included task descriptions and example explanations, and the answers required the model to output in JSON format. Then the LLaMA2-13B model was selected as the base model, and the few-shot strategy was used for SFT fine-tuning, ultimately developing a model capable of quickly extracting the names of each author and their actual contribution information. After testing, the final model achieved an accuracy of 95.1% on the test set. It is important to note that for certain fixed referential terms like 'the author,' this study employed a database matching method to replace them with the names of all the authors in the paper, effectively ensuring data validity.

Finally, a DOL classification mapping model was developed. Considering the need for standardized and large-scale statistical analysis of author contribution paragraphs, this study relied on the 14 categories of the CRediT taxonomy and the LLaMA2-13B base model to train a large model capable of quickly performing DOL classification. For training data construction, 200 author contribution paragraphs were manually labeled with their corresponding DOL categories. To quickly supplement the training set and reduce manual labeling costs, GPT-3.5 was used to automatically generate an additional 800 related entries, totaling 1,000 entries. During training, an 80:20 train/test ratio was applied, and the model was successfully trained to construct a chain structure linking 'author name-author contribution-DOL roles' (CRediT categories).

In our study, the input efforts of the target papers align with the five types of scientific contributions. To calculate the input efforts of the target papers, we established a mapping between 14 DOL roles and the 5 scientific contribution types (as shown in Table 4). This mapping is primarily based on the official annotations of CRediT and the relevance of each role to the contribution type. Notably, only the '5. Investigation' role is mapped to both 'Experimental' and 'Data-based' contributions because this role explicitly includes 'performing the experiments' and 'data/evidence collection'. Given the effort expended by individuals assuming multiple roles, we employed fractional counting to calculate the credit score for each type of role in the target paper, with the calculation method shown in Formula 1.

$$C_l = \sum_{a \in A_l} \frac{1}{D_a} \ , \ A_l \subseteq \{a_1, a_2, \ldots, a_n\}$$

*$C_l$ represents the credit score for a specific role type, $D_a$ represents the number of roles the author a assumes, $A_l$ denotes the set of authors with role l in the paper, and n represents the number of authors in the paper.*

Based on the credit scores and the mapping relationships of each role, we further calculated the distribution of input efforts of DOL, with the calculation method shown in Formula 2.

$$P_i = \sum_{l \in L_i} \frac{C_l}{n} \ , \ L_i \subseteq \{l_1, l_2, \ldots, l_{14}\}$$

$P_i$ represents the proportion of a specific input effort of DOL，$C_l$ represents the credit score for a specific role type, n is the number of authors in the paper, and $L_i$ is the set of roles corresponding to input effort i.

**Table 4** Mapping relationship between input efforts of DOL and DOL

| Types of input efforts of DOL | Type of division of labor |
|---|---|
| Theoretical | 1. Conceptualization, 13. Writing-Original Draft Preparation, 14. Writing-Review & Editing |
| Methodological | 6. Methodology, 9. Software |
| Experimental | 3. Formal Analysis, 5. Investigation, 11. Validation, 12. Visualization |
| Data-based | 2. Data Curation, 5. Investigation, 8. Resources |
| Other | 4. Funding Acquisition, 7. Project Administration, 10. Supervision |

*Feature Analysis of the paper's Scientific Contributions*

*Measurement of the relationship between the actual contributions and the input efforts of DOL.* The first step in feature analysis is to measures the correlation between the actual contribution distribution and the input contribution distribution of papers. We used the Pearson correlation coefficient to reflect this relationship. In terms of calculation, we employed both aggregate correlation statistics for the entire collection of papers and paper-level correlation statistics. This multi-faceted approach to correlation calculation ensures the robustness of the results.

*Measurement of the internal relationship among different types of scientific contributions from Input-output perspectives.* The next step is to explore whether there is a coupling or isolation relationship between the various scientific contribution types of the paper. Our method involves first analyzing the likelihood of multiple types of scientific contributions occurring simultaneously. For example, we examine the proportion of papers where authors are involved in multiple types of input efforts of DOL from the input perspective, and the proportion of papers producing multiple types of actual contributions from the output perspective. Subsequently, we construct a co-occurrence matrix of scientific contributions for samples where multiple types of contributions are present, and calculate the co-occurrence strength between each pair of contribution types using diagonal normalization methods. Regarding the co-occurrence frequency statistics, in the input perspective, we add 1 to the co-occurrence frequency between different types of input efforts of DOL for each author involved in

multiple types of input work. In the output perspective, we first determine the primary type of output contribution for each paper based on its proportion. Then, we compare the proportions of other types with the main type. If the difference is less than 15%, it is considered that the paper has also produced a certain degree of other types of contributions in addition to the main one.

## 4. Results

**4.1 Distributions of the actual contributions of papers**

(1) Overall distributions of the actual contributions of papers

First, we explored the actual contribution distribution of papers in our experimental set (Table 5). It shows that experimental contributions account for the highest proportion (56.51%) among the papers measured, followed by theoretical and methodological contributions (20.34% and 14.14%, respectively). Other types of contributions have the lowest proportion, about 1.11%. This result indicates that most papers primarily influence and advance scientific research through experimental work and empirical results. In contrast, theoretical and methodological innovations are relatively challenging; these contributions help the research community establish and improve research foundations and methodologies, providing support and guidance for empirical research. However, proposing new theories and methods has a higher threshold compared to empirical research, leading to fewer outputs and, consequently, lower contribution levels. Notably, data-based contributions are the lowest among all types, revealing that despite the current data-driven scientific paradigm, the trend of reusing scientific data remains weak. This finding aligns with research related to scientific data citations (Park et al., 2018; Zhao et al., 2018). In the future, it is crucial to enhance the visibility and usability of scientific data in papers by implementing data management and usage policies and promoting best practices for open sharing.

**Table 5** The number and proportion of citing content corresponding to various types of output contributions

| Contribution type | Citing content number | Citing content number% |
|---|---|---|
| Theoretical | 312,084 | 20.34% |
| Methodological | 216,933 | 14.14% |
| Experimental | 867,159 | 56.51% |
| Data-based | 121,284 | 7.90% |
| Other | 17,007 | 1.11% |

(2) Distributions of the actual contributions of papers among different disciplines

Next, we analyzed the differences in the actual contribution distribution of papers across disciplines. A past study by Porter et al. (1988) identified the types of scientific contributions through manual recognition of representative papers from different fields and found that various disciplines have distinct focuses in their scientific contributions. To verify this conclusion, we compared the selected fields of chemistry, physics, and neuroscience from previous study with their statistical results and extracted papers from these disciplines in our experimental set to calculate their actual contribution proportions. As shown in Table 6, although experimental contributions are the highest across the three fields (consistent with the overall distribution), methodological contributions are more prominent in chemistry, theoretical contributions dominate in physics, and experimental contributions are more evident in neuroscience. This result aligns with the conclusions of the previous study. The differences in scientific contribution emphasis across disciplines may be influenced by factors such as research objects, methods, goals, and historical traditions specific to each discipline.

**Table 6** Distributions of the actual contributions of papers in three disciplines

| Disciplines | The actual contributions | | | | |
| --- | --- | --- | --- | --- | --- |
| | Theoretical | Methodological | Experimental | Data-based | Other |
| chemistry | 27.83% | **32.16%** | 36.57% | 1.59% | 1.84% |
| physics | **30.84%** | 17.08% | 48.79% | 1.95% | 1.34% |
| neurosciences | 18.39% | 15.21% | **58.28%** | 7.49% | 0.62% |

**4.2 Relationships between the actual contributions and the input efforts of DOL**

(1) Comparison between the overall distribution of the actual contributions and the input efforts of DOL

The study further explored which input effort factors might relate to the actual contributions of papers. We first calculated each paper's input efforts of DOL, and then, to visualize the input-output relationship in scientific contributions, we calculated the total proportion of various contributions from both input and output perspectives in our dataset (Figure 2). The results revealed that while the input efforts of DOL focused more on theoretical work, the actual contributions emphasized experimental contributions. Furthermore, we calculated the overall correlation coefficient (r=-0.0389,

p=0.1041) and paper-level correlation coefficient (r=0.3923, p=0.4528) between the distribution of input efforts of DOL and the actual contributions, finding no significant correlation between labor input patterns and the actual output contributions trends. Considering the varying importance of different types of scientific contributions, we also implemented several weighting schemes and calculated correlation coefficients, which similarly showed no significant relationships (details in Table S2 & S3).

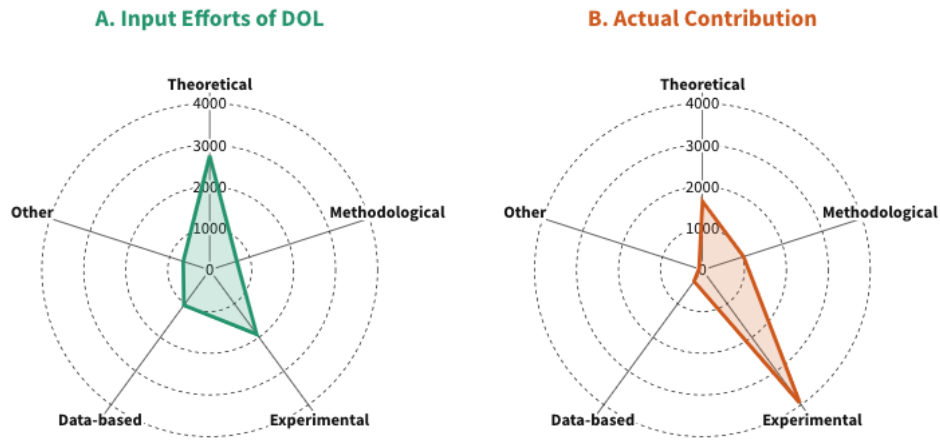

**Figure 2.** (A) Total proportion of each output contribution type in sample papers. (B) Total proportion of each input efforts type of DOL in sample papers.

(2) Input-output relationship of each type of scientific contribution

Next, we delved into the input-output relationship specific to each scientific contribution type. Although the overall input efforts distribution appears unrelated to the actual contribution trends, specific labor inputs might still influence a particular type of actual contribution. To investigate this, we proposed a dual perspective approach for analyzing the input-output relationship of each contribution type. From the input perspective, we categorized papers by the dominant labor type (5 types) based on the highest proportion of input efforts of DOL, and then examined the actual contribution distribution within these groups. For the output perspective, we grouped papers by the dominant actual contribution type (5 types) and examined the input efforts of DOL distribution in these groups.

The analysis revealed no significant relationship between each input effort type and actual contribution type. From the input perspective, regardless of the primary labor input effort type, the highest actual contributions were experimental, followed by theoretical and methodological. From the output perspective, regardless of the primary actual contribution type, the highest input efforts of DOL types were theoretical, followed by experimental and data-based (Figure S2). This aligns with the total proportion distribution of scientific contributions observed in Figure 2. Considering the possible impact of numerical differences in proportions, we normalized the average proportions by dividing each scientific contribution type's input efforts of DOL by the average proportions from the entire paper set. The same method applied to the output perspective.

The normalized results in Figure 3 show that, from the input perspective, greater input in a specific contribution type corresponds to higher actual output in that type. A similar pattern emerged from the output perspective: greater actual output in a contribution type correlates with higher labor input. Additionally, high input in one type often corresponds to high output in multiple contribution types, and vice versa. For example, when papers have higher input in data-based contributions, the output in both data-based and methodological contributions exceeds baseline levels. Similarly, higher experimental labor input correlates with above-average output in experimental and data-based contributions.

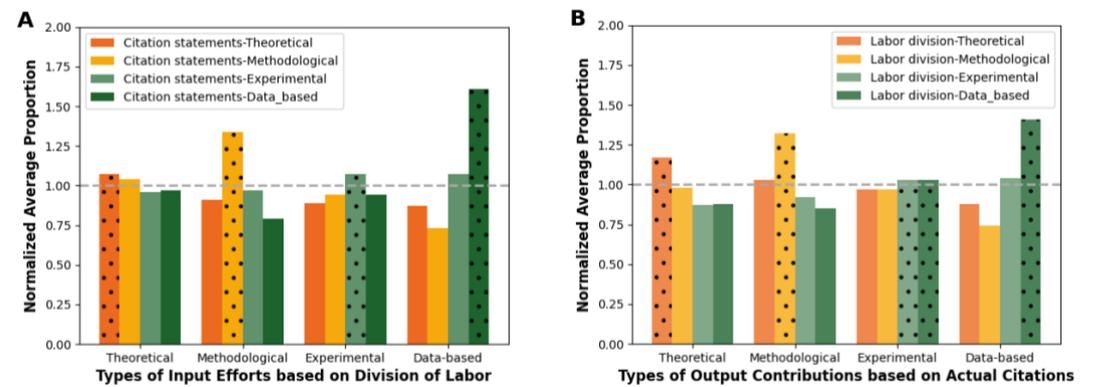

**Figure 3.** (A) Normalized average proportion distribution of output contributions for each type of input efforts of DOL (when this type of effort constitutes the majority). (B) Normalized average proportion distribution of input efforts of DOL for each type of output contributions (when this type of output contributions constitutes the majority). Note that the type of other contribution is not shown in the input and output analysis since this type has no clear contribution intention.

To investigate the specific DOL types corresponding to actual contributions, we analyzed the relationship between each actual contribution and the 14 DOL roles. Figure 4 shows the normalized average proportions of these roles when a paper's proportion of a specific contribution type is highest. It reveals that high theoretical contributions correspond to increased labor in software, conceptualization, supervision, and writing-original draft preparation. High methodological contributions relate to increased labor in software, validation, methodology, and visualization. High experimental contributions relate to greater labor in investigation, resources, and formal analysis. High data-based contributions involve more labor in data curation, software, and visualization. We performed a similar analysis from the output perspective, with results in Appendix Figure S3.

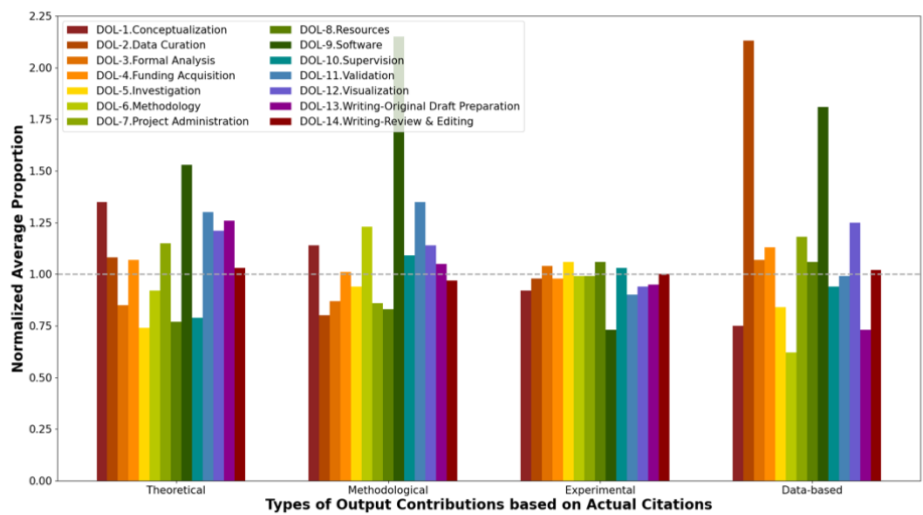

**Figure 4.** Normalized average proportion distribution of DOL roles for each type of actual contributions output (when this type of actual contributions output constitutes the majority).

## 4.3 Internal relationships among different types of scientific contributions from Input-output perspectives

Scientific research is a complex ecosystem where different types of scientific contributions may either depend on each other or operate independently. Therefore, we analyze the coupling and isolation relationships between various types of scientific contributions from both the input (i.e., the input efforts of DOL) and output (i.e., the actual contributions) perspectives, in order to gain a deeper understanding of the diversity of scientific elements' relationships.

Our analysis of the proportion of multiple types of scientific contributions co-occurring in the sample set reveals that in the input perspective, the proportion of co-occurrence of multiple types of scientific contributions is 78.1%, while in the output perspective, it is only 19.78%. This indicates that during the research exploration phase, various input efforts of DOL are more likely to co-occur, with researchers often playing versatile roles. However, when research results are published and enter the dissemination stage, the actual contributions of the papers are often more focused and less frequently encompassing multiple types.

Further analysis of the co-occurrence strength between contribution types in cases where multiple types are present (Figure 5) shows that, from the input perspective, theoretical contributions exhibit the strongest co-occurrence with other types, indicating that those involved in theoretical work are often versatile across different labor division types. Conversely, methodological contributions have the weakest co-occurrence with other types, suggesting that those engaged in methodological research tend to work independently. Among the contribution types, the co-occurrence strength between theoretical and experimental contributions is the strongest (0.7549), indicating that these types of work are often performed by the same individuals, while the co-occurrence strength between data-based and methodological contributions is the weakest (0.3895), suggesting that these types of work are more likely to be done separately.

From the output perspective, the analysis shows that experimental contributions have the strongest co-occurrence with other types of contributions, demonstrating that strong experimental contributions are more likely to lead to other types of contributions. Data-based contributions, on the other hand, exhibit the weakest coupling with other types, showing that data contributions are relatively independent. Among the output contributions, theoretical contributions and experimental contributions have the strongest co-occurrence (0.7818), which implies that these types of contributions often occur together, while data-based and theoretical contributions have the weakest co-occurrence (0.1512), revealing a lower correlation between these types.

Overall, regardless of the perspective, there is a stronger co-occurrence between theoretical and experimental contributions, reflecting the close integration of theoretical foundations and practical applications in scientific research activities.

Additionally, we analyzed the internal relationships between various types of scientific contributions across different disciplines (see Appendix Figure S4). The results indicate that the internal relationship characteristics of scientific contributions across the three disciplines are consistent with the overall findings. Furthermore, we found that in the input perspective, the co-occurrence strength of each type of input efforts of DOL in the field of chemistry is slightly higher than in physics and neuroscience, which may indicate a higher degree of collaboration and cross-discipline interaction in chemical research. In the output perspective, each type of output contribution in the field of neuroscience shows moderately higher co-occurrence strength with other types compared to chemistry and physics, which may be attributed to the highly interdisciplinary nature and comprehensive application value of neuroscience research.

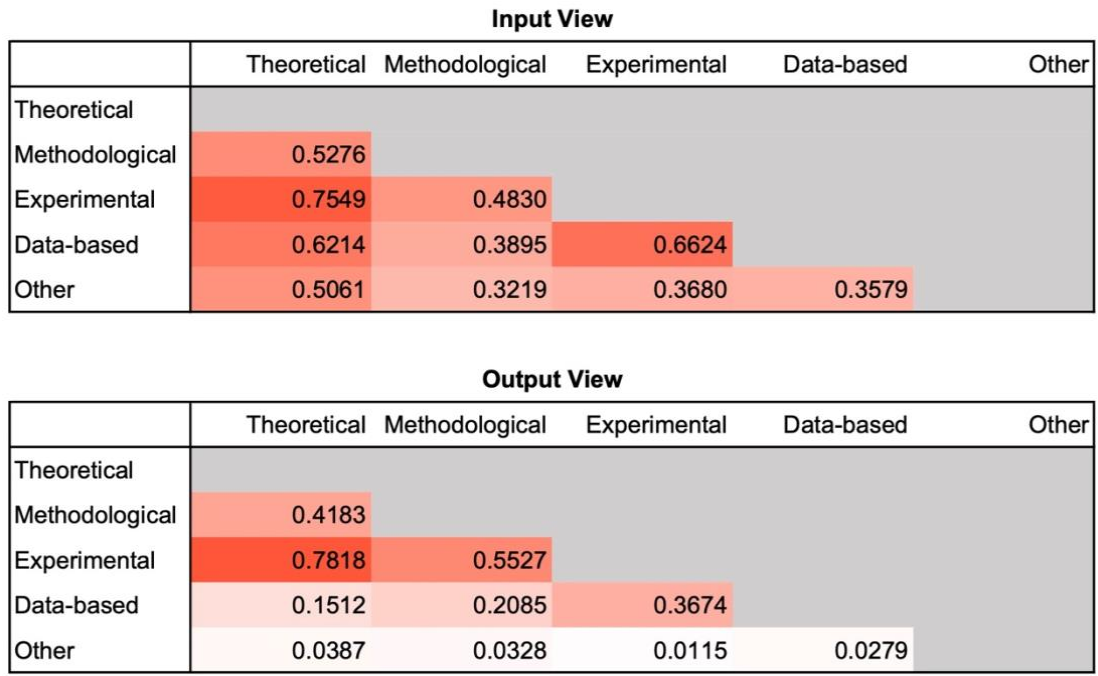

Figure 5. Co-occurrence strength between each type of scientific contribution (The statistical results include both input and output perspectives).

## 5. Discussion and Conclusion

This study proposes a method for measuring scientific contribution types based on citation context and explores the characteristics of scientific contributions in 7,041 papers from the journals *Nature* and *Science* from the perspectives of citation context and labor division. Overall, our results indicate that the citation context in citing papers can reveal the actual impact of target papers on the scientific community, while labor division reflects the investment patterns of papers in various contributions from the dimensions of resource allocation and collaborative behavior. Each perspective of measuring scientific contributions has its emphasis.

(1) Characteristics of the paper's actual contributions
Our study shows that papers make more concentrated experimental contributions to other research during the publication and dissemination stages. This peer-recognized contribution tendency differs from previous trends based on authors' self-perceived contributions, which are mainly theoretical and methodological (Chao et al., 2023; Porter et al., 1988). This discrepancy may arise from differences in the purposes of authors and peers in academic writing and citation behavior. Authors often emphasize their innovative ideas and methodological developments when writing papers, whereas experimental results are more easily verified and repeated by peers and have a more direct impact on scientific progress, making them more likely to be cited. This feature is important for optimizing scientific evaluation methods. For example, matching the self-identified contribution points with those extracted from the citation context can help identify research achievements with theoretical and methodological innovations more clearly. We also found that the actual contribution patterns vary across different disciplines, consistent with existing qualitative research conclusions. This variation is likely related to the research goals and historical development of the disciplines. For instance, physics research often involves fundamental laws and phenomena of nature, requiring theoretical models and mathematical derivations, thus establishing a strong theoretical research foundation. The field of chemistry focuses on developing and improving synthesis and analytical methods, with experimental techniques playing a crucial role in both basic research and practical applications in industry, materials, and medicine. Neuroscience relies on experimental techniques (such as neuroimaging, behavioral experiments, and electrophysiological recordings) to obtain data and validate theoretical hypotheses, making experimental research predominant.

(2) Relationship between input efforts of DOL and actual contributions
Our findings reveal no significant correlation between the overall labor input patterns and the actual output trends in scientific contributions. However, there is a positive correlation between input and output in specific scientific contribution types, reflecting the "more work, more gain" effect. For instance, a paper aiming to make a theoretical contribution should have a significantly higher-than-average input in theoretical labor division, and the same applies to other types. Furthermore, the input and output patterns of scientific contributions also exhibit a "synergistic effect."
A detailed analysis reveals that the overall input patterns and output trends are influenced by factors such as disciplinary specialization and the cross-synergistic effects among different contributions. Different disciplines or fields have specific labor input patterns and emphasize different contribution types (as verified by us, with physics focusing on theoretical contributions, chemistry on methodological contributions, etc.). This professional specialization and diversity in scientific practice complicate the overall input-output relationship, making it difficult to observe clear associations. Different scientific contribution types within a paper may influence and collaborate with each other; for example, methodological labor input may improve the quality of experiments and data, leading to corresponding contribution outputs. This cross-synergistic effect further obscures the correlation between overall input and output. However, when analyzing each scientific contribution type, consistency is observed in labor input and contribution output. This indicates that to achieve above-average contributions of a certain type, the corresponding labor input must also be above average. This can be achieved by having specialists focus on one research task or allocating more personnel to handle it, reflecting the professionalization of scientific research that leads to the "more work, more gain" effect. Additionally, we observed a synergistic effect between labor input and actual output in specific scientific contribution types, where high input in one type of labor division results in higher contributions not only in that type but also in other related contributions, corroborating our earlier judgment of the complexity in the overall correlation.

(3) Internal relationships between scientific contribution types

Our findings indicate that the internal relationships between various types of scientific contributions exhibit different interaction patterns from the input and output perspectives. During the exploratory phases of scientific activities, research teams often consist of generalists who can perform multiple tasks (Haeussler & Sauermann, 2020). However, once the research results enter the dissemination and communication phase, the actual contributions of papers tend to be more focused, often displaying only one type of contribution rather than multiple. This phenomenon reflects both the rich collaborative and labor resource allocation models in scientific activities and, to some extent, the high specificity of scientific issues and the constraints imposed by technological resources on research practices.

Specifically, from the perspective of input coupling, it is observed that theoretical contributors are more likely to engage in multiple types of labor divisions, providing guidance and leadership across various research activities while establishing the theoretical foundation of research problems. In contrast, those engaged in methodological research tend to be more independent, primarily offering methodological support to other tasks but less involved in other types of work. From the perspective of output associations, research with a focus on experimental contributions is more likely to simultaneously generate other types of scientific contributions. Conversely, papers with data-based contributions exhibit stronger independence, reflecting the unique methods and research objectives of data-driven studies. Furthermore, from both the input and output perspectives, there is a strong co-occurrence between theoretical and experimental contributions, which substantiates the widely accepted principle in scientific research that "theory and practice must be closely integrated". The theoretical contributions in papers provide explanations and predictions of natural phenomena through logical reasoning and abstract models, while the empirical applications validate the correctness and feasibility of theories through experiments. These two types of contributions are interdependent, collectively advancing knowledge accumulation and innovation.

*Limitations*

Our study has some limitations. First, we used two authoritative comprehensive journals to measure scientific contributions, focusing on natural sciences with a relatively limited sample size. Additionally, scientific contributions are complex issues, involving not only labor division input and actual citation output but also potentially related to funding input, novelty of research questions, and other factors. Future research could explore multi-factor causal analyses of contribution input and output.

*Future Research Directions*

Future research can delve deeper into the following directions. First, expand the journal scope by collecting more citation context data and author contribution statements from various scientific fields to conduct broader analyses of scientific contribution characteristics. It could also include comparative analyses between natural sciences and humanities and social sciences. The current study focuses on the most prominent research achievements in *Nature* and *Science* regarding scientific contribution patterns. However, is the scientific contribution pattern the same for ordinary papers? This question merits investigation. Second, compare the actual contribution types of papers with authors' self-perceived contributions and peer recognition to uncover differences in behavioral patterns between actual effects, self-perception, and peer recognition of scientific contributions. Besides the actual contributions reflected in citing papers' citation context studied here, self-perception can be measured through traditional author statements, while peer-recognized contributions can be revealed through expert reviews. The peer review content during the review stage of a paper can directly reflect peers' perceptions of the current research achievements. Journals such as *PLOS* openly provide peer review content, allowing further measurement of a paper's actual contributions based on such texts. The three different scientific contributions represent the three stages of scientific contribution generation, and comparative analysis can deepen the understanding of the essence of scientific contributions. Additionally, drawing on existing study, further exploration of paper value evaluation methods through scientific contribution identification can enhance the rationality and effectiveness of scientific evaluation systems.

## Acknowledge

This work was supported by National Social Science Fund of China (NSSFC, Grant No. 23BTQ103) and China Association for Science and Technology Foundation (Grant No. 2023QNRC001). We thank the anonymous reviewers as well as the academic editor of the paper for their numerous comments and suggestions that helped to improve the paper.